# Investigating the Electromechanical Behavior of Unconventionally Ferroelectric Hf$_{0.5}$Zr$_{0.5}$O$_2$-based Capacitors Through Operando Nanobeam X-ray Diffraction

**Authors:** Evgenios Stylianidis, Pranav Surabhi, Ruben Hamming-Green, Mart Salverda, Yingfen Wei, Arjan Burema, Sylvia Matzen, Tamalika Banerjee, Alexander Björling, Binayak Mukherjee, Sangita Dutta, Hugo Aramberri, Jorge Íñiguez, Beatriz Noheda, Dina Carbone*, Pavan Nukala*

**Affiliations:**

**E. Stylianidis**

Department of Physics and Astronomy, University College London, Gower Street, WC1E 6BT, London, United Kingdom

**P. Surabhi, P. Nukala**

Center for Nanoscience and Engineering, Indian Institute of Science, Bengaluru, 560012, India

**R. Hamming-Green, M. Salverda, A. Burema, T. Banerjee, B. Noheda**

Zernike Institute for Advanced Materials, University of Groningen, Groningen, 9747 AG, The Netherlands

**Y. Wei**

Frontier Institute of Chip and System, Fudan University, Shanghai, China

**A. Björling, D. Carbone**

MAX IV Laboratory, Lund University, PO Box: 118, SE- 221 00, Lund, Sweden

**S. Matzen**

Center for Nanoscience and Nanotechnology, Paris-Saclay University, CNRS, 91120, Palaiseau, France

**B. Mukherjee, S. Dutta, H. Aramberri, J. Íñiguez**

Materials Research and Technology Department, Luxembourg Institute of Science and Technology (LIST), L-4362 Esch/Alzette, Luxembourg

**J. Íñiguez**

Department of Physics and Materials Science, University of Luxembourg, L-4422 Belvaux, Luxembourg

**B. Noheda**

CogniGron center, University of Groningen, Groningen, 9747 AG, The Netherlands

Corresponding authors*:

Pavan Nukala, email: pnukala@iisc.ac.in

Dina Carbone, email: gerardina.carbone@maxiv.lu.se

**Keywords**

**Abstract**

Understanding various aspects of ferroelectricity in hafnia-based nanomaterials is of vital importance for the development of future non-volatile memory and logic devices. Here, the unconventional and weak electromechanical response of epitaxial La$_{0.67}$Sr$_{0.33}$MnO$_3$/Hf$_{0.5}$Zr$_{0.5}$O$_2$/La$_{0.67}$Sr$_{0.33}$MnO$_3$ ferroelectric capacitors is investigated, via the sensitivity offered by nanobeam X-ray diffraction experiments during application of electrical bias. It is shown that the pristine rhombohedral phase exhibits a negative linear piezoelectric effect with piezoelectric coefficient ($d_{33}$) ~ -0.5 to -0.8 pmV$^{-1}$. First-principles calculations support an intrinsic negative piezoresponse. In addition, it is found that the piezoelectric response is suppressed above the coercive voltage. For higher voltages, and with the onset of DC conductivity throughout the capacitor, a second-order effect is observed. The electromechanical response observed in this work is clearly different from that of normal ferroelectrics, again underlining the unconventional nature of polarization switching in the samples.

## 1. Introduction

For more than a decade now, hafnia-based nanomaterials are justifiably among the most studied materials in the field of ferroelectrics and microelectronics, both from a fundamental and an application perspective.[1,2] Hafnia and its compounds show a unique type of ferroelectricity that becomes robust at the nanoscale,[3] unlike traditional perovskite ferroelectrics that suffer from depolarization effects at reduced dimensions.[4] The discovery of ferroelectricity in hafnia, combined with its silicon compatibility, initiated a new wave of studies for novel ferroelectric memories and logic nanodevices.[5,6]

Understanding the stabilization of a ground state ferroelectricity at the nanoscale, as well as the physical mechanism behind ferroelectric switching in this peculiar ferroelectric, has proved far from trivial. The most usually reported phase of ferroelectric hafnia is a metastable polar orthorhombic (*o*-) phase,[2] while a polar rhombohedral (*r*-) phase has been stabilized under certain conditions when La$_{0.7}$Sr$_{0.3}$MnO$_3$ (LSMO)-buffered Hf$_{0.5}$Zr$_{0.5}$O$_2$ (HZO) is grown on SrTiO$_3$ (STO) substrates.[7] Interestingly, as demonstrated by integrated differential phase contrast scanning transmission electron microscopy (iDPC STEM),)[8] this *r*-phase lacks the polar/non-polar sublayer structure that has been reported to be responsible for the flat-phonon band and, thus, for the dipolar localization that possibly gives rise to unconventional, non-cooperative, switching behavior of ferroelectric hafnia.[9] Furthermore, the remnant polarization in this phase arising from the polar nature of the *R3m* phase, which itself is stabilized by oxygen deficiency, is < 9 µCcm$^{-2}$.[9] Nevertheless, macroscopic ferroelectric characterization on 5 nm thick *r*-phase samples showed very large switching currents corresponding to having a remnant polarization of ~35 µCcm$^{-2}$, comparable to or larger than those observed in *o*-phase hafnia.[10] This discrepancy has been resolved through in situ electron microscopy experiments, that clearly showed that polarization switching is intertwined with oxygen vacancy migration, and is different from normal ferroelectrics.[8,11]

Conventional ferroelectric switching is accompanied by an electromechanical response, characterized by butterfly-like strain *vs* electric field loops. Understanding the electromechanical response of unconventional hafnia-based ferroelectrics can give more insights into the origins of ferroelectricity. However, despite the massive interest in the ferroelectric properties of hafnia-based materials, the explorations on electromechanical behavior are still in the nascent stages. Electrostrain measurements on thick polycrystalline La:HfO$_2$ films reported by Schenk et al. show a classic ferroelectric butterfly loop, with the upturns in the strain curve occurring at the coercive field, and a positive piezoelectric coefficient *d$_{33}$* ~7.7 pmV$^{-1}$.[12] Recent theoretical and experimental works reported that a negative longitudinal piezoelectric coefficient is possible in the orthorhombic phase of HfO$_2$, a behavior that is usually found in organic ferroelectrics.[13–15] More interestingly, first-principles simulations predict that the sign of the piezoelectric coefficient *d$_{33}$* can be reversed by tuning the local environment without affecting the global polar *o*-phase symmetry or switching its polarization.[15]

Point defects and defect-migration-induced giant electrostriction have recently gained tremendous attention in sister compounds of hafnia such as doped ceria and bismuth oxide.[16,17] The switching characteristics of *r*-phase hafnia have similar defect-aided origins.[8] Thus, it is paramount to understand the electromechanical behavior of these systems, and enquire further into the origin of the observed switching currents in hafnia-based systems.

In this report, we present results from synchrotron diffraction experiments during application of external bias on La$_{0.67}$Sr$_{0.33}$MnO$_3$/Hf$_{0.5}$Zr$_{0.5}$O$_2$/La$_{0.67}$Sr$_{0.33}$MnO$_3$ capacitors grown epitaxially on (001)-oriented SrTiO$_3$ substrates. By following the evolution of the diffraction peaks of our materials hand-in-hand with a bursting voltage scheme, we probe the electrostrain behavior of the thin films at various applied fields. We show that *r*-phase HZO exhibits only a negative linear piezoresponse at voltages even larger than the coercive voltage. A second-order occurs at much higher voltages where leakage or DC conductivity is significant. Our analysis shows that the negative *d$_{33}$* is intrinsic for the weakly-polar *r*-phase, a result supported by our first-principles simulations. We ascribe the second-order response to effects such as Joule heating and defect migration.

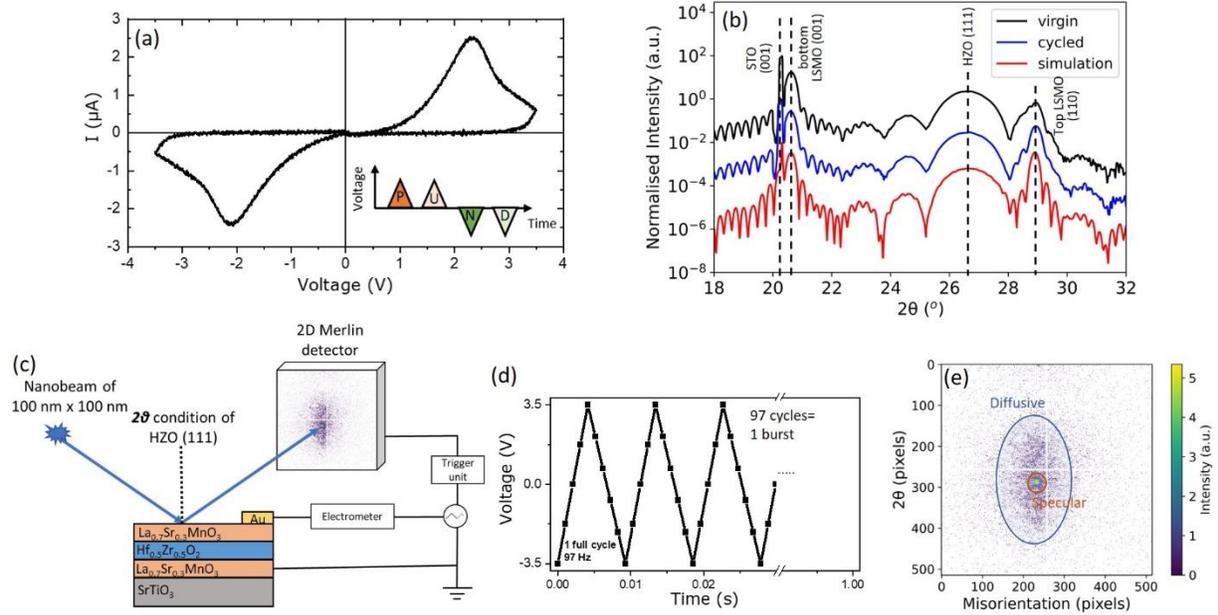

**Figure 1: (a)** PUND-corrected macroscopic electrical characterization performed offline, using a 1 kHz triangular voltage pulses scheme (inset). **(b)** Specular X-ray diffractograms of the device under test at the virgin and cycled states, as well as a simulated curve. The curves are offset for clarity. **(c)** Schematic illustration of the experimental setup used during the operando synchrotron experiment. **(d)** Triangular voltage pulses with 97 Hz are used. Black squares represent the times where X-ray frames are acquired. **(e)** Snapshot of typical data on the 2-dimensional detector: the diffuse and the high-intensity specular part of the HZO (111) Bragg peak are highlighted.

## 2. Results and Discussion

Electrical pre-characterization and crystalline coherence

LSMO/HZO/LSMO capacitors compatible with operando biasing experiments were fabricated using photolithography-based techniques (see Methods). Figure 1(a) depicts the positive up / negative down (PUND)-corrected current response as a function of the applied voltage, showing the characteristic polarization-switching peaks, as previously identified for this system.[7]

Figure 1(b) shows representative X-ray diffraction (XRD) specular diffractograms before (virgin) and after (cycled) the synchrotron experiment, including several cycles of voltage biasing. The XRD spectra show clearly a well-defined (001) Bragg peak from the bottom LSMO layer at $2\theta=20.76°$, the (111) peak from the HZO layer at $2\theta=26.60°$, and the (110) peak from the top LSMO layer at $2\theta=28.96°$. The presence of well-defined intensity oscillations (Laue oscillations) reveals the highly crystalline coherence of the system. Oscillations with two different periods can be observed: the short period around the (001) LSMO Bragg peak is relative to a thickness of around 35 nm, while the larger period originates from the 5 nm thick HZO (111) layer. The top layer of LSMO (also 35 nm thick) is highly oriented along the pseudo-cubic (p.c.) direction

<110>$_{p.c.}$. A good simulation of the diffractograms was obtained using a model proposed by Lichtensteiger,[18] based on the kinematical theory. The as-deposited samples show a crystallographic polarization pointing towards the top LSMO layer (or "polarization up", see reference [8]). All the subsequent experiments begin by resetting to the "polarization up" state.

Operando diffraction experiment

A diffraction experiment while applying an electrical bias was performed at the diffraction endstation of the NanoMAX beamline of MAX IV Laboratory in Lund, using a 100 nm diameter beam and a wavelength of 1.379 Å. The setup of the experiment is graphically illustrated in figure 1(c). Triangular voltage waveforms varying from -$V_{max}$ to $V_{max}$ of 97 Hz were applied in 50-60 successive bursts, always grounding the bottom electrode. Each burst lasted 1 second (97 triangular voltage pulses, see figure 1(d)), and structural data was acquired every millisecond on a 2D detector aligned at $2\theta$=26.60º, which is the (111) Bragg peak of HZO. In this setup, any HZO (111) lattice parameter variation translates into a subtle movement of the Bragg peak along the vertical axis on the 2D detector. The instantaneous displacement current was simultaneously measured using an electrometer capable of detecting nA currents.

The signal on the detector image contains two contributions (figure 1(e)): a sharp central peak, which arises from the crystal truncation rod of the coherent heterostructure, and a broad diffuse intensity (along the $2\theta$ axis) that is generated by the HZO film and relates to the lattice spacing of the HZO (111) planes ($d_{111}$). We followed in-situ the evolution of the diffuse intensity during the application of the triangular voltage wave, by acquiring data with a frequency about 10 times higher than the voltage pulses. The 97 Hz frequency was chosen to reduce the occurrence of artifacts due to the sole acquisition scheme. The measurement of repeated bursts was used to improve the signal-to-noise ratio by integrating all the images corresponding to the same voltage (this is practically done by associating to each dataset a specific time-stamp over all the bursts). The collection of such large statistics, combined with the intrinsically high strain resolution provided by the X-ray diffraction, allowed us to determine the variation of the lattice parameter by tracking the changes of center of mass (COM) of the (111) peak with the applied bias, to a precision of <0.01 pm. The use of a focused beam was necessary to locate the μm size device on the large sample support and to check heterogeneity of behavior in the device. Owing to the sample being aligned at the (111) Bragg peak, the effective size of the beam on the inclined sample was 400 nm along the direction of propagation (vertical, in the geometry used).

Representative results from the operando experiments are presented in figure 2(a) and (b) for two selected voltage amplitudes. For the case of a voltage amplitude of 2 V, we observe a small linear change of the lattice constant, in the opposite direction with respect to the applied field: it expands when a negative field is applied, and contracts when the field is reversed, giving rise to a negative effective piezoelectric coefficient $d_{33}$ (see figure S1 in Supplementary). Fourier Transform (FT) analysis of the lattice parameter variation over time for a repeated number of bursts, shows a single peak at 97 Hz, corresponding to the drive voltage frequency, confirming that the XRD effectively captures a lattice variation caused by the applied voltage. We note that the linear behavior with a negative slope (corresponding to a negative $d_{33}$) is observed for all voltage amplitudes below and up to 2 V (figure S2 in Supplementary); however, at 2.8 V, the peak at 97 Hz in the FT disappears, indicating almost no change in the HZO (111) lattice parameter with voltage. These results are uniform across the device area, reproducible, and do not depend on the sample history.

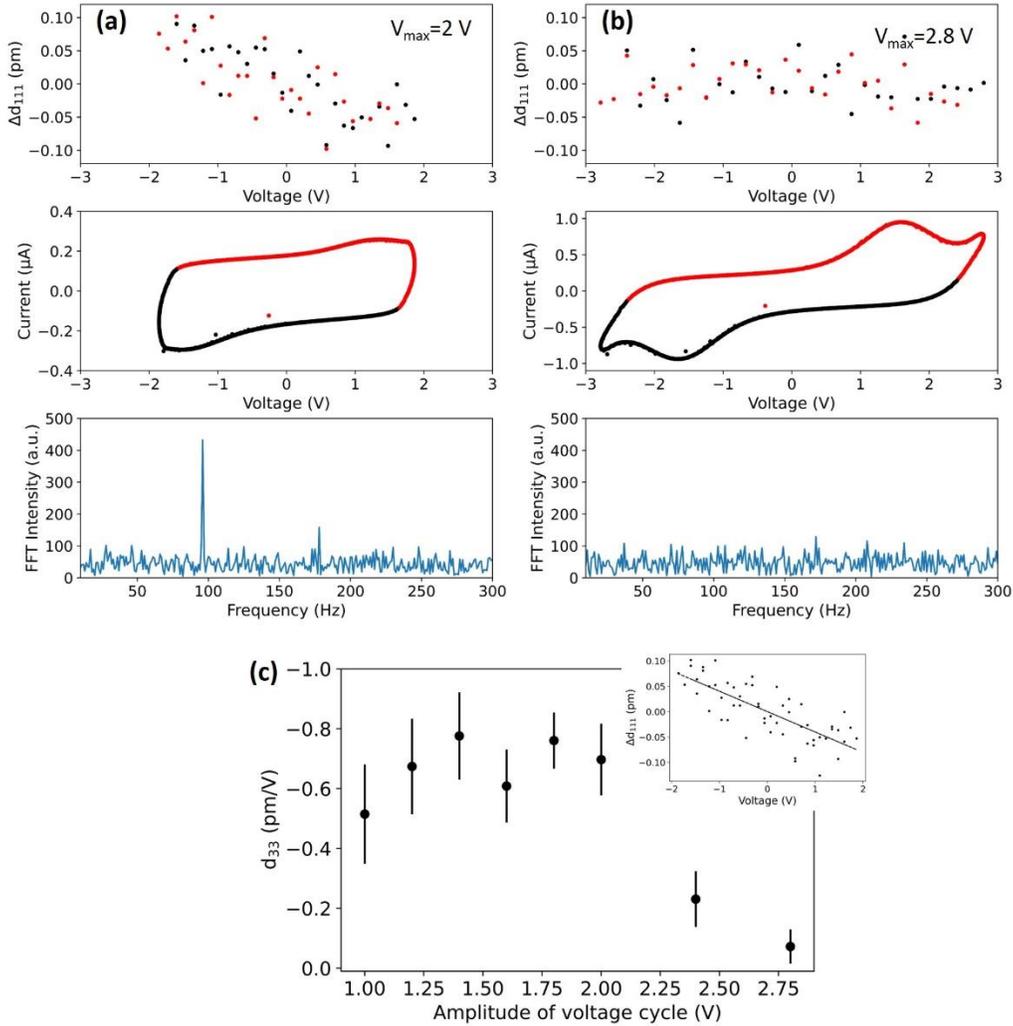

**Figure 2:** Measurements of the unit-cell strain performed by applying an amplitude of **(a)** 2 V and **(b)** 2.8 V. Top: variation of lattice parameters of HZO as extracted from the shift in reciprocal space of the HZO (111) reflection. Middle: the I-V curves as measured by the electrometer during the diffraction experiment. Red and black points/lines represent the data collected from positive and negative going voltage cycles, respectively. Bottom: Fourier transform of the temporal evolution of the HZO (111) COM over the whole measurement. **(c)** The longitudinal piezoelectric coefficients $d_{33}$ as calculated from linear fits of the lattice parameter response to each of the applied voltages (inset) as a function of the amplitude of the applied voltage.

To quantify the changes in the lattice parameter of HZO, we performed a linear fit of the lattice constant *vs* voltage curves and estimated $d_{33}$ at various maximum voltages as shown in figure 2(c). Within the fitting error, we observe that $d_{33}$ does not change significantly up to 2 V and ranges between -0.5 pmV$^{-1}$ to -0.8 pmV$^{-1}$. At 2.4 V, $d_{33}$ decreases to -0.3 pmV$^{-1}$ and becomes zero at 2.8 V. It is important to note that the observed reduction in $d_{33}$ for $V_{max}$>2 V occurs when the current switching peak starts becoming apparent in the displacement current, as observed in the *I-V* characteristics which were simultaneously measured during the burst experiment (figure 2(a) and (b)). Note that, when $V_{max}$ is in above the coercive voltage, in the range 2-2.8 V, the device does not show conventional ferroelectric butterfly type strain response, rather still shows a simple linear weak piezoelectric response.

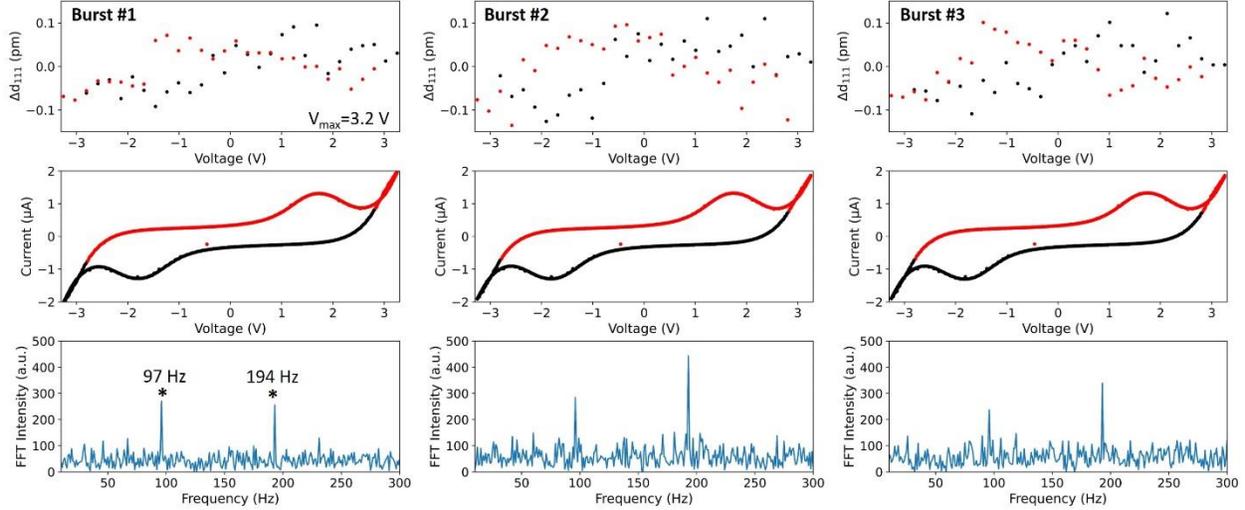

**Figure 3:** Repeated unit-cell strain measurements with voltage amplitude of 3.2 V. Top: the variation of HZO (111) out-of-plane lattice parameter. Middle: the instantaneous *I-V* response during the experiment. Bottom: the calculated Fourier transform of the lattice parameter variation.

For even higher voltages (>2.8 V), the electromechanical response of HZO shows both a first and a significant second-order behavior (figure 3), as clearly revealed by the appearance of two peaks, at 97 Hz and at 194 Hz in the Fourier transform. Repeated burst measurements at the same voltage show the repeatability of this response. It is important to note that the onset of this second order behavior is uncorrelated from the ferroelectric switching. This correlates, however, with the leakage behavior at large fields ($V_{max}$=3.2 V and above) and increases with increasing leakage. We confirmed this by measuring the electromechanical response of damaged highly leaky capacitors which still show the 194 Hz peak in the Fourier transform (figure 3 in Supplementary).

Discussion-Model and physical behavior

Our results suggest two distinct regimes of electromechanical response in our devices. At low fields, we observe a negative linear response, consistent with a weak piezoelectric response of a weakly polar *r*-phase. However, as the amplitude of the applied bias increases, the electromechanical response decreases and it is strongly reduced when the applied bias is large enough to complete the polarization switching process. Surprisingly, for applied voltages higher than the coercive voltage, only a small negative linear behavior is observed (i.e. figure S2 for an applied bias of 2.4 V). This clearly shows that the electromechanical response is unlike conventional ferroelectrics, which show the onset of second order behavior corresponding to the domain nucleation and growth about the coercive field. The second order response observed in our samples occurs at much larger voltages ($V_{max}$>2.8 V), where leakage becomes significant. Leakage-induced second order strain effects include Joule heating and non-classical electrostriction (through ion migration) and these mechanisms can explain the high-field behavior of our devices.[16,19]

We use first-principles techniques to compute the intrinsic piezoresponse of the *r*-phase HZO (see Methods). We obtain a value of $d_{33}$=-7.6 pmV$^{-1}$ for a strain-state that has a positive polarization of 1.01 µCcm$^{-2}$ along the rhombohedral axis. The selected strain state corresponds to the defect-free *R3m* phase, with the lattice parameters measured experimentally. Our earlier electron microscopy studies indicate that, our samples indeed are stabilized in a low-strain *R3m* phase albeit with oxygen vacancies. Without considering the presence of oxygen vacancies the *R3m* phase has a very low crystal polarization, as calculated from DFT.[7] Inclusion of oxygen vacancies allowed us to estimate the crystal polarization values ~9 µCcm$^{-2}$.

The full elastic and piezoelectric tensors computed from first-principles here are thus on defect-free *R3m* phase, and are provided in Supplementary table S1, S2 and S3. The mechanism behind the development of a spontaneous polarization in the *R3m* phase is quite convoluted (involving several modes and an improper origin), and ultimately leads to a relatively small polarization. Therefore, it is impossible to come up with an intuitive physical picture as the one obtained for the *o*-phase in ref. [15], since here the negative sign of the longitudinal piezoresponse arises from the (partial) compensation of several weakly-polar distortions. At any rate, our calculations do yield an intrinsic anomalous longitudinal piezoresponse in agreement with the experimental observation. The computed effect is about 7 times larger than the measured one, and it should be considered an upper limit corresponding to an ideal (infinite, periodic) film where size effects are not considered.

Additionally, let us note that it is unclear how to compute the extrinsic contribution to the piezoresponse from atomistic simulations; this would crucially depend on the particular extrinsic mechanism (such as the migration of oxygen vacancies) considered, and falls beyond the scope of this work.

An intrinsic negative piezoelectric response has been previously observed in hafnia-based compounds, the nature of which still remains contradictive. On the one hand, Chouprik et al.[20] found that an anomalous negative piezoelectric response can arise locally in domains of polycrystalline *o*-phase HZO films, due to internal fields created by accumulation of oxygen vacancies at the metal-HZO interfaces. However, the absence of any built-in bias in the electrical measurements suggest that this effect is rather marginal in our system. On the other hand, using first-principles calculations, Dutta and collaborators[15] showed that intrinsic negative piezoelectricity is possible in hafnia-based compounds due to the peculiar local environment of the active oxygen atoms in the orthorhombic phase. In the same work, experimental results based on piezoresponse force microscopy revealed negative $d_{33}$ values in the same order of magnitude (3 times larger than ours) in *o*-phase polycrystalline samples. Along similar lines, our experimental and first-principles results lead us to propose that the *R3m* phase has also intrinsically negative piezoelectric coefficients.

This work shows that the electromechanical response is clearly decoupled from ferroelectric switching, supporting unconventional origins of ferroelectricity in these samples. Our previous experiments[8] revealed that polarization switching is intertwined with oxygen vacancies migration, and this work lends further credence to this mechanism. Thus, any effect related to ionic migration cannot be neglected from the electromechanical response in the high voltage regime, above the coercive field. For instance, defects and ionic migration can cause non-classical electrostriction, as found in sister fluorite compounds of hafnia such as Gd-doped $CeO_2$ [16,19] and (Y, Nb)-stabilized $Bi_2O_3$.[17].

3. Conclusions

We investigated the electromechanical behavior of *r*-phase HZO via operando synchrotron studies using high-flux nanofocused X-ray beam diffraction. At low fields, our results revealed a negative linear piezoresponse with $d_{33}$ ~ -0.5 pmV$^{-1}$ to -0.8 pmV$^{-1}$, which is related to intrinsic negative piezoelectricity in rhombohedral HZO, as supported by first-principles calculations performed in low strain, defect-free *R3m* phase. Even above the coercive voltage, we still observe only first order response, albeit weakened further, and the absence of a conventional ferroelectric butterfly loop. We also show that leakage directly correlates with the second order strain response (at even high voltages) with possible origins in Joule heating or ion migration. Our findings clearly lend credence to the previous works that show that ferroelectricity in *r*-phase HZO is unconventional and driven by and intertwined to oxygen vacancy migration.

4. Experimental methods

Growth

The materials $La_{0.7}Sr_{0.3}MnO_3$ (30 nm)/$Hf_{0.5}Zr_{0.5}O_2$ (6 nm)/$La_{0.7}Sr_{0.3}MnO_3$ (30 nm) were grown on (001)-oriented $SrTiO_3$ substrates by pulsed laser deposition (PLD) as described elsewhere.[7]

Device fabrication

The fabrication flow diagram of the devices can be found elsewhere.[8] Square capacitors with an area of 50 μm x 50 μm were patterned using standard photolithography and Ar-ion milling. The capacitors were electrically connected to Ti/Au contacting pads, while their surface was kept optically clear to be probed with the nanobeam. The contacting pads were then Au wire bonded to chip carriers.

Macroscopic electrical characterization pre-synchrotron

The capacitors were electrically tested in the lab prior the synchrotron experiment using the standard PUND (positive-up-negative-down) measurement scheme (inset of figure 1(d)). Triangular pulses of 1 kHz in frequency and 3.5 V in amplitude were used, with a delay time of 250 μs in between each pulse. The data presented in figure 1 (a) are the PUND-corrected data: the current response during the up (down) pulse was subtracted from the current response of the positive (negative) pulse, resulting in a current response that contains only the displacement current. A Keithley 4200A-SCS parameter analyzer equipped with Source Measure Units (SMUs) and a Pulse Measurement Unit (PMU) was used for the measurement.

In-situ biasing burst diffraction experiment

The operando experiments were performed at the NanoMAX beamline of the MAX IV Laboratory in Lund, Sweden.[21] A beam of wavelength 1.379 Å and a beam size of 100 nm in diameter was used. We used a burst voltage scheme consisting of triangular pulses of 97 Hz frequency while acquiring X-ray images and current measurements at 1 kHz. A R&S HMF2550 pulse generator was used for the formation of the biasing train, and an electrometer (ALBA) was used to measure the current, while the bias was sampled with a PandABox (Quantum Detectors) acquisition unit. A 2-dimensional Merlin detector (Quantum Detectors) was used, which was at a fixed position during the bursting experiment, corresponding to the *2θ* angle of the main Bragg peaks of our films.

First-principles calculations

We performed density functional theory (DFT) calculations as implemented in the Vienna Ab initio Simulation Package (VASP).[22,23] For the exchange-correlation functional we used the Perdew-Burke-Ernzerhof formulation for solids (PBEsol)[24] of the generalized gradient approximation (GGA). The atomic cores were treated within the projector-augmented wave approach[25], considering the following states explicitly: 5p, 6s and 5d for Hf; 4s, 4p, 4d and 5s for Zr; and 2s and 2p for O. We used an 800 eV energy cut-off for the plane-wave basis set. Brillouin zone integrals were computed in a 4x4x1 Gamma-centered k-point grid corresponding to a simulation supercell of 72 atoms. We arranged the hafnium and zirconium atoms in alternating layers along the [111] rhombohedral direction in order to preserve the *R3m* symmetry of the *r*-phase. The $d_{111}$ lattice spacing was fixed to the experimentally observed value of 2.99 Å, and the structure was allowed to relax under this constraint until residual forces fell below 0.01 eVÅ$^{-1}$. The piezoresponse tensor was computed using density functional perturbation theory (DFPT)[26] as readily implemented in VASP. The elastic tensor (which we used to obtain the piezoelectric strain tensor *d* from the DFPT-computed piezoelectric stress tensor *e*) was obtained using a finite-differences scheme, with each degree of freedom perturbed by 0.015 Å. The polarization was computed using the Berry phase approach within the modern theory of polarization.[27]

5. Acknowledgments

We acknowledge MAX IV Laboratory for time on Beamline NanoMAX under Proposal 20190954. Research conducted at MAX IV, a Swedish national user facility, is supported by the Swedish Research council under contract 2018-07152, the Swedish Governmental Agency for Innovation Systems under contract 2018-04969, and Formas under contract 2019-02496. Work at LIST was supported by the Luxembourg National Research Fund though grants INTER/NOW/20/15079143/TRICOLOR (B.M., H.A., J.I.) and PRIDE/15/10935404/MASSENA (S.D.).

6. Conflict of Interest

The authors declare no conflict of interest.

## 7. Data availability Statement

Data that support this study are available upon reasonable request by the corresponding author.

**Supplementary information**

**Figure S1**

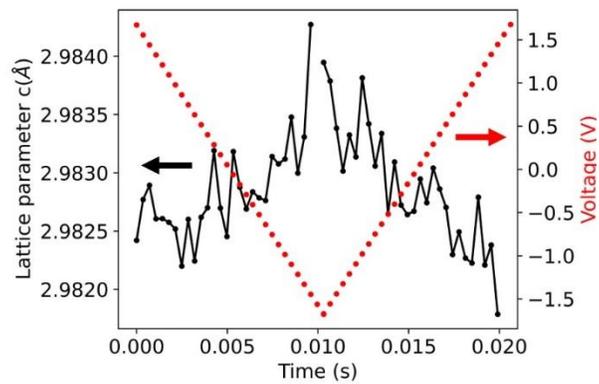

**Figure S1:** Modulations of the center-of-mass of the (111) HZO out-of-plane lattice parameter and the applied voltage as a function of time, for an averaged 97 Hz voltage pulse.

**Figure S2**

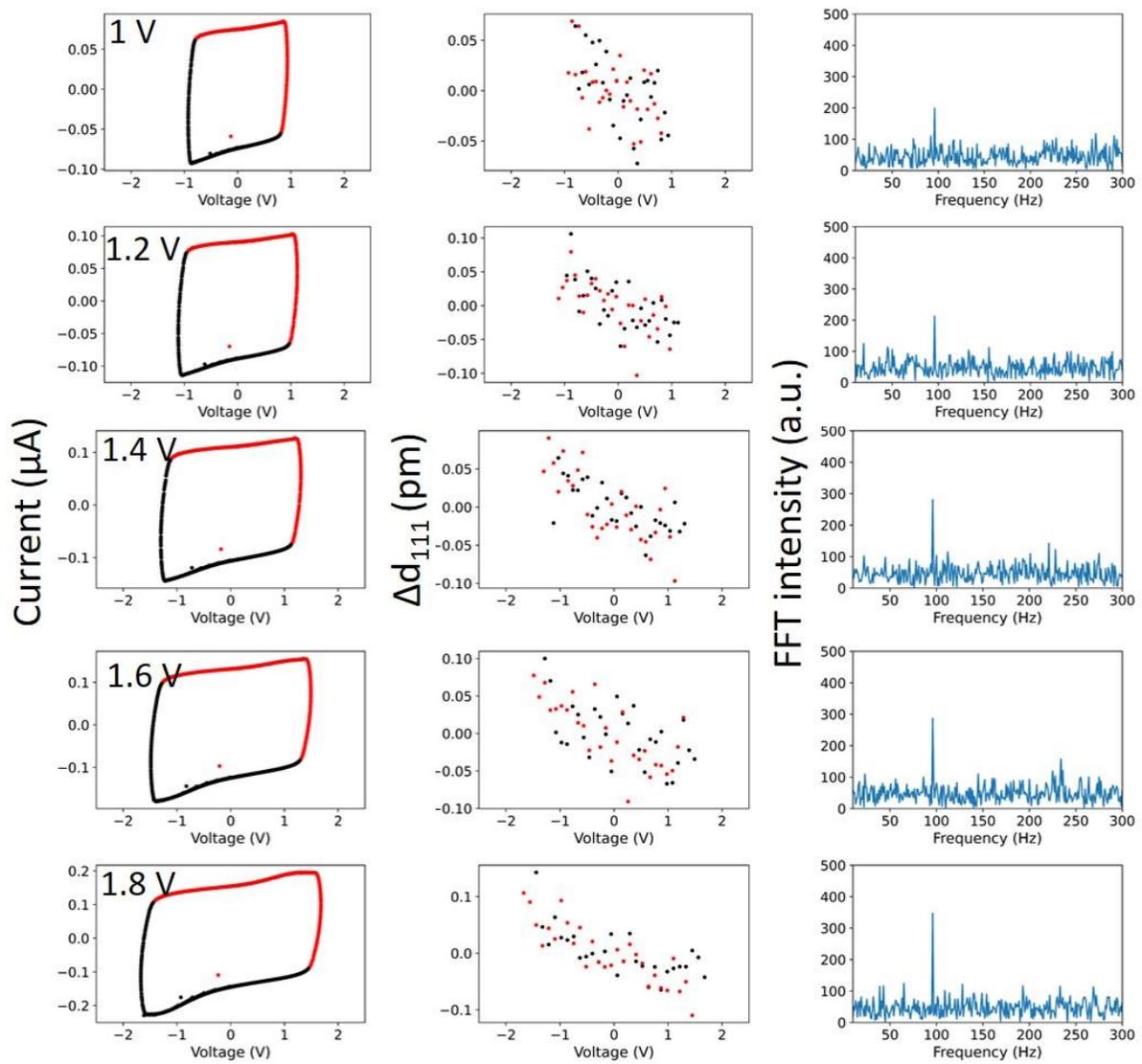

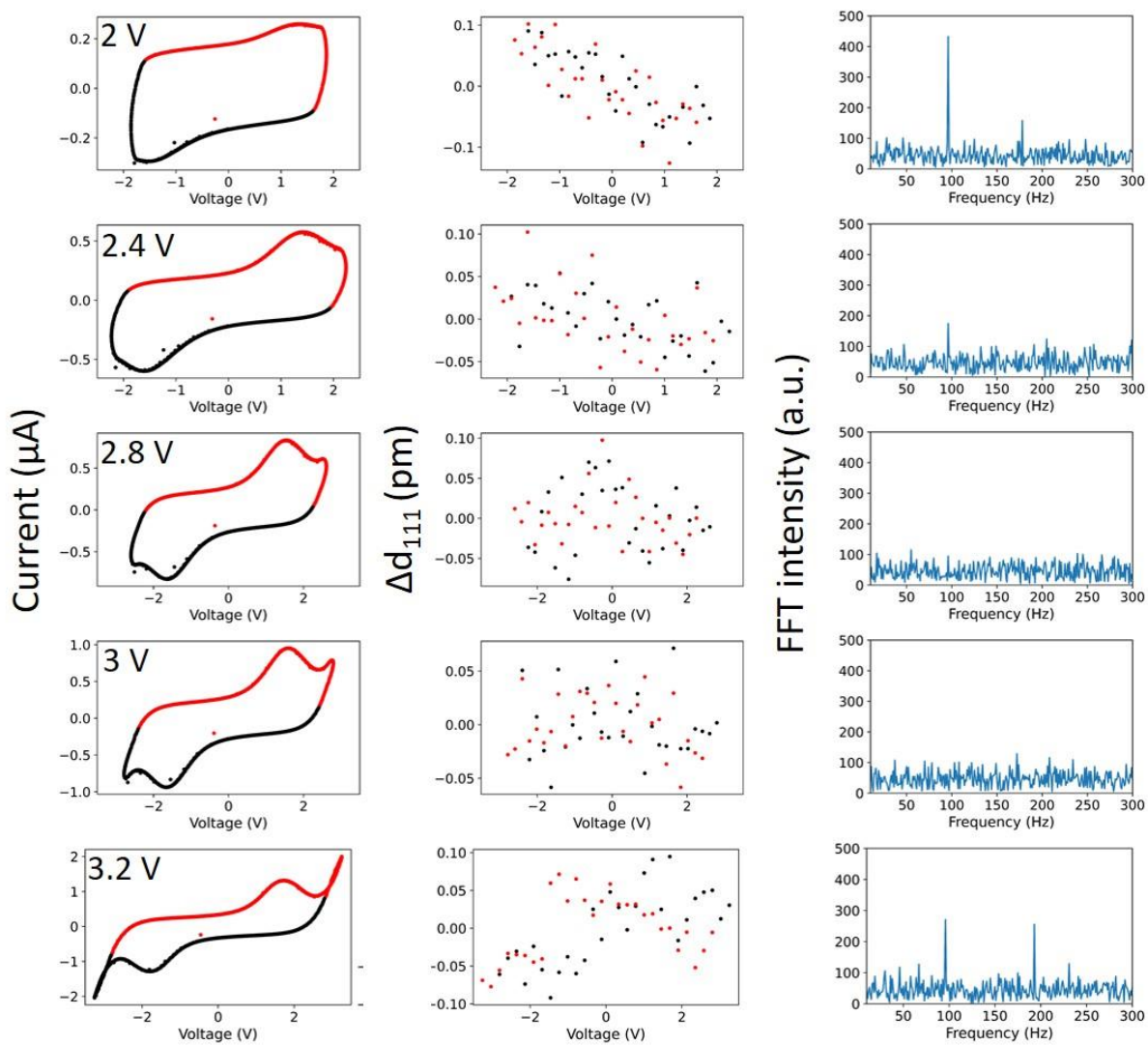

**Figure S2:** Summarized results of our operando synchrotron studies on the electromechanical response of rhombohedral HZO. **Left column:** online I-V characteristics measured simultaneously with the acquisition of X-ray data. **Middle column:** relative modulations of the HZO (111) lattice parameter as a function of the applied voltage. **Right column:** Fourier transform of the temporal evolution of the HZO (111) lattice parameter.

**Figure S3**

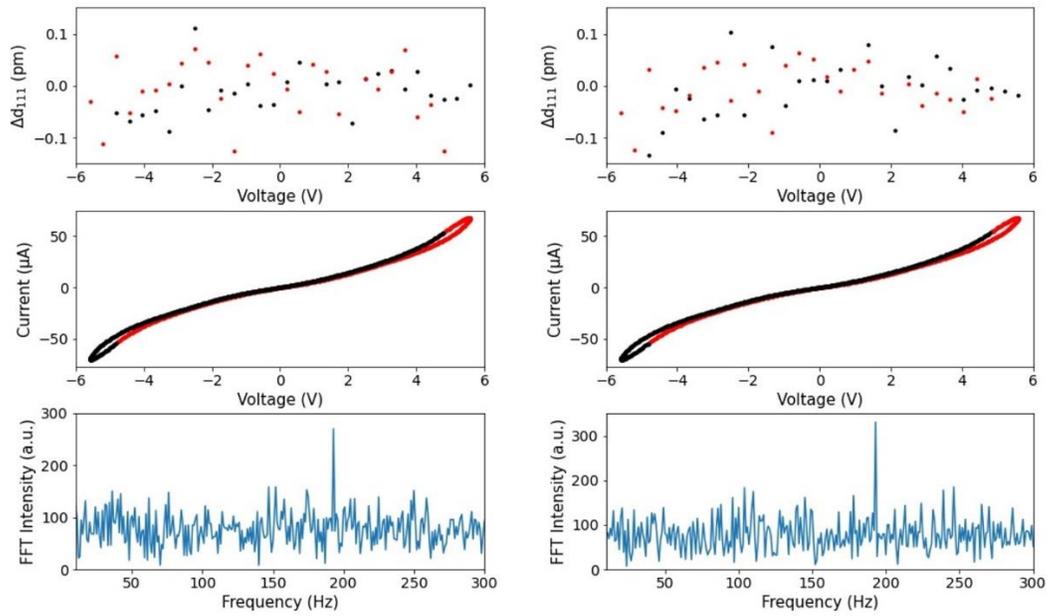

**Figure S3:** Electromechanical response of HZO in a highly leaky capacitor. A peak at 194 Hz is observed in the Fourier transform which suggests that the second order response has extrinsic origins rather than coupled to intrinsic polarization switching process in the HZO film.

**Supplementary Table S1:** Computed direct piezoelectric tensor ($e$) in Cm$^{-2}$.

$$e = \begin{bmatrix} 0.0 & 0.0 & 0.0 & -0.0533 & 0.0 & -0.0265 \\ -0.0533 & 0.0533 & 0.0 & 0.0 & -0.0267 & 0.0 \\ 0.2447 & 0.2447 & -0.6412 & 0.0 & 0.0 & 0.0 \end{bmatrix}$$

**Supplementary Table S2:** Computed elastic tensor ($C$) in GPa.

$$C = \begin{bmatrix} 361.05 & 145.42 & 189.46 & 0.0 & 60.27 & 0.0 \\ 145.42 & 361.05 & 189.46 & 0.0 & -60.27 & 0.0 \\ 189.46 & 189.46 & 249.68 & 0.0 & 0.0 & 0.0 \\ 0.0 & 0.0 & 0.0 & 107.81 & 0.0 & 60.27 \\ 60.27 & -60.27 & 0.0 & 0.0 & 150.98 & 0.0 \\ 0.0 & 0.0 & 0.0 & 60.27 & 0.0 & 150.98 \end{bmatrix}$$

**Supplementary Table S3:** Calculated converse piezoresponse tensor ($d$) in pmV$^{-1}$.

$$d = \begin{bmatrix} 0.0 & 0.0 & 0.0 & -0.51 & 0.0 & 0.03 \\ -0.25 & 0.25 & 0.0 & 0.0 & 0.03 & 0.0 \\ 3.34 & 3.34 & -7.64 & 0.0 & 0.0 & 0.0 \end{bmatrix}$$